\documentclass[journal=nanolett,manuscript=article]{achemso}

\usepackage{chemformula} 
\usepackage[T1]{fontenc} 
\usepackage{xcolor}
\usepackage{physics}
\usepackage{graphicx}
\usepackage{bm}
\usepackage{amsmath}
\usepackage{natbib}
\usepackage{multirow}

\usepackage{sectsty}
\SectionNumbersOn

\title{Remote control of spin polarization of topological corner states}

\author{Yinong Zhou}
\affiliation{Department of Physics and Astronomy, University of California, Irvine, California 92697-4575, USA}

\author{Ruqian Wu}%
\email{wur@uci.edu}
\affiliation{Department of Physics and Astronomy, University of California, Irvine, California 92697-4575, USA}
\begin{document}
	
\vspace{80pt}


\clearpage

\begin{abstract}
In two-dimensional higher-order topological insulators, the corner states are separated by a non-negligible distance. The crystalline symmetries protect the robustness of their corner states with long-range entanglement, which are robust against time-reversal breaking perturbations. Here, we demonstrate the possibility of direct control of the topological corner states by introducing the spin degree of freedom in a rhombus-shaped Kekul\'{e} nanostructure with local magnetization and local electric potential. By applying a local magnetization on one corner, the other corner can also be strongly spin polarized. By further applying a local electric potential at the same corner, the sign of the spin polarization can be reversed at both corners. We demonstrate the material realization in a $\gamma$-graphyne nanostructure with Mn adsorption and Si replacement at one corner by using the first-principles calculations. Our studies give a showcase of the remote correlation of quantum states in higher-order topological materials for spintronic and quantum applications.
\end{abstract}

\clearpage

Research interest in topological insulators (TIs) \cite{hasan2010colloquium} has stayed extremely high as their gapless edge or surface states are robust against perturbations and hence have potential for diverse technological innovations \cite{jia2012defects,hu2019topologically,ni2020robustness,zhang2021superior}. Recently, the proposal of electric multipole insulators \cite{benalcazar2017quantized} inspired explorations of a new class of topological materials, so-called higher-order topological insulators (HOTIs) \cite{zhang2013surface,peng2017boundary,langbehn2017reflection,song2017d,benalcazar2017electric,schindler2018higher,schindler2018higher2}, which show lower-dimensional topological corner or hinge states. Several two-dimensional (2D) lattices have been proposed for the realization of the HOTI phase, including square lattice \cite{benalcazar2017quantized,peng2017boundary,langbehn2017reflection,song2017d,benalcazar2017electric,schindler2018higher}, breathing Kagome lattice \cite{ezawa2018higher,bolens2019topological}, Kekul\'{e} lattice \cite{wu2016topological,liu2017pseudospins,liu2019two,liu2019helical,mizoguchi2019higher,zangeneh2019nonlinear,lee2020fractional}, and non-periodic quasicrystals \cite{huang2021generic}. The difference between the intercell and intracell hopping parameters drives these systems into the higher-order topological phase by opening a gap for the edge states and generating more localized topological corner states. Unlike the conventional bulk-edge correspondences \cite{bernevig2013topological} in 2D TIs, the higher-order corner states manifest the bulk-corner correspondence  \cite{hasan2010colloquium,qi2011topological} or the edge-corner correspondence \cite{ezawa2020edge,khalaf2021boundary}. As the corner states are protected by certain crystalline symmetries, the higher-order topology is preserved in the presence of disorders \cite{proctor2020robustness,coutant2020robustness,xie2021optimization} even with the time-reversal symmetry breaking \cite{hu2021disorder}.

It is known that the topological edge states in 2D TIs lead to the quantum spin Hall effect (QSHE) \cite{kane2005quantum,bernevig2006quantum,konig2007quantum} or the quantum anomalous Hall effect (QAHE) \cite{haldane1988model,liu2008quantum,chang2013experimental,laughlin1983anomalous}, and their possible applications in quantum computing and spintronic devices have been extensively investigated  \cite{nayak2008non,yokoyama2009giant,zhang2010quantum,breunig2021opportunities}. Comparatively, the potential use of the corner states in 2D HOTIs has rarely been discussed because even their realization in real materials remains as a challenge. Experimental studies of HOTIs to date were mostly based on metamaterials \cite{serra2018observation,peterson2018quantized,noh2018topological,imhof2018topolectrical,xue2019acoustic,ni2019observation,zhang2019second,zhang2020low}, and some new phenomena have been reported such as the valley-selectivity of corner states of a sonic crystal \cite{zhang2021valley}. In this regard, it is essential to find a real solid HOTI, establish a profound understanding, and design conceptual devices for benefiting from the discovery of this new topological phase. Since spin is one of the most important factors in all topological studies \cite{haldane1988model,liu2008quantum,chang2013experimental,laughlin1983anomalous}, it is natural to perceive that the control of the spin degree of freedom of the topological corner states is crucial for the application of HOTIs. Although the artificial pseudospin degree of freedom has been introduced for the topological corner state \cite{liu2019helical}, the study of their real-spin degree of freedom is still barren.

In this letter, we propose a possibility of making remote magnetoelectric control in rhombus-shaped Kekul\'{e} nanostructure, through theoretical studies using the tight-binding (TB) method and the computational studies using the first-principles calculations with tunable local magnetization ($M$) and local electric potential ($V$). By magnetizing one corner, the opposite corner separated by a non-negligible distance can also be spin-polarized. Furthermore, because of the unique correspondence of the topological corner states, electric manipulation at the same corner may alter the sign of spin polarization (SP) at both corners. Importantly, we demonstrate the material realization of HOTI using a $\gamma$-graphyne nanostructure with the Density Functional Theory (DFT) calculations which show identical responses to the manipulations of $M$ and $V$ as predicted by the TB model. Here, $M$ and $V$ are applied by using Mn adsorption and Si substitution at one corner of the $\gamma$-graphyne flake. This study extends the understanding of HOTI and gives a showcase of using the topological corner states in quantum and spintronic devices.

 We start from the TB model Hamiltonian for the Kekul\'{e} lattice:
 \begin{equation}
 	H_0 = t_0\sum_{\langle{ij}\rangle}{c_{i}^\dagger}c_{j} + t_1\sum_{\langle{i'j'}\rangle}{c_{i'}^\dagger}c_{j'}.
 \end{equation}
 Here, $c_{i}^\dagger$ is the operator of the electron creation on-site $i$. $\langle{ij}\rangle$ and $\langle{i'j'}\rangle$ represent the nearest-neighbor hopping for intracell ($t_0$) and intercell ($t_1$), respectively, as shown in the inset of Figure 1a. As reported in previous studies \cite{wu2016topological,liu2017pseudospins,liu2019two,liu2019helical,mizoguchi2019higher,zangeneh2019nonlinear,lee2020fractional}, a HOTI phase is created on the Kekul\'{e} lattice when $t_1>t_0$. Here, we choose $t_1/t_0=1.25$ in the following discussions. The gapped edge states and parity calculations confirmed the high-order topology with $Z_2=1$ (see Supporting Information Section I). In order to show the topological corner states, we construct a rhombus-shaped nanostructure with an 11$\times$11 Kekul\'{e} lattice (Figure 1a). There is a pair of states located in the gap of the bulk and edge states around the Fermi level, representing the bonding and antibonding corner states (Figure 1b). The wavefunctions of these corner states are evenly distributed at two corners, as shown in Figure 1c,d. Note that the topological local states only appear at the 120$^{\circ}$ corners due to the chiral charge cancellation at the 60$^{\circ}$ corners for the bipartite lattices \cite{liu2019two}.
 
 \begin{figure}[t]
 	\centering
 	\includegraphics[width=0.7\textwidth]{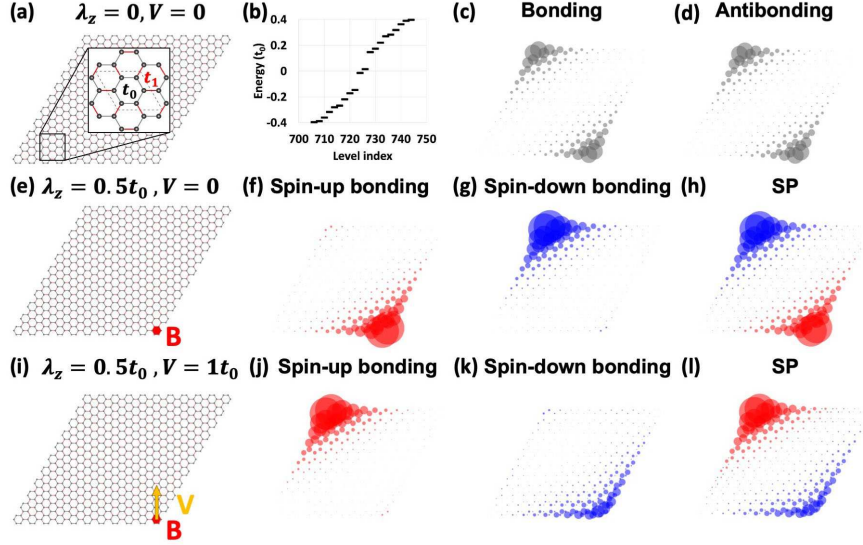}
 	\caption{\label{fig:epsart} The rhombus-shaped Kekul\'{e} nanostructures (a-d) without $M$ and $V$, (e-h) with $M$ (red hexagon) at the bottom-right corner, and (i-l) with $M$ and $V$ (yellow arrow) at the same corner ($V>\lambda_z$). (a,e,i) Schematic figure of the nanostructures. The inset in (a) shows the Kekul\'{e} unit cell with intracell hopping $t_0$ (grey bonds) and intercell hopping $t_1$ (red bonds). (b) The discrete energy levels of (a). (c,d) Wavefunction distributions of bonding and antibonding corner states in (b) with spin degeneracy. (f,g,j,k) Wavefunction distributions of spin-up and spin-down bonding corner states (f,g) with $M$ and (j,k) with $M$ and $V$. (h,l) Spin polarization distributions of the bonding corner states (f,g) and (j,k), respectively. The size of the grey, red, and blue circles is proportional to the local charge density of corner states.}
 \end{figure}
 
 Next, we consider the interplay between $M$ and $V$ on the corner states. The total Hamiltonian becomes:
 \begin{equation}
 	H_{tot} = H_0 - \lambda_z\sum_{\alpha}{c_{\alpha}^\dagger}(\sigma_z\otimes\tau)c_{\alpha} + V\sum_{\alpha}{c_{\alpha}^\dagger}(I_2\otimes\tau)c_{\alpha},
 \end{equation}
 where $\lambda_z$ represents exchange splitting, $V$ represents electric potential, $\alpha$ is the spin index. $\sigma_z$ is the Pauli matrix and $I_2$ stands for a 2$\times$2 identity matrix. $\tau$ represents 2$\times$2 matrix acting on the two corners, where $\tau=\left(\begin{matrix}1&0\\0&0\end{matrix}\right)$ represents $M$ or $V$ applied on the bottom-right corner (Figure 1e,i).  The SP of the corner states is defined as $SP = {|\braket{\alpha_\uparrow}{\psi}|}^2-{|\braket{\alpha_\downarrow}{\psi}|}^2$, where $\psi$ is the wavefunction of the corner states. We first choose parameters $\lambda_z=0.5t_0$, and $V=t_0$ to illustrate the influence of $M$ and $V$.
 
 When the bottom-right corner is magnetized, as shown in Figure 1e, the spin degeneracy is broken for the corner states. The wavefunction of the bonding corner state becomes unevenly distributed at two corners, as shown in Figure 1f,g. The spin-up $M$ applied on the bottom-right corner induces a spin-down polarization at the top-left corner spontaneously. The SP shows antiparallel couplings for the two corners (Figure 1h). Interestingly, when $V$ ($V>\lambda_z$) is applied on the same corner, as shown in Figure 1i, the spin-up bonding corner state becomes localized at the top-left corner (Figure 1j), which is due to the newly generated spin-down corner state localized at the bottom-right corner (Figure 1k). As a result, the sign of the SP is reversed at both corners, comparing Figure 1h and 1l. Consequently, the manipulation of $M$ and $V$ at one corner can not only induce the SP at the other corner but also can reverse the sign of the SP at both corners. This offers an innovative way to remotely control the magnetization of HOTI flakes.
 
 To understand these important findings, we investigate the evolvement of the corner states with $M$ and $V$, respectively. First, let’s consider the change of $\lambda_z$ when $V=0$, as shown in Figure 2a. When $M$ is applied, the spin degeneracy is broken for the corner states. The four corner states are located at the corner tips, as the wavefunction distributions shown in Figure 2a for $\lambda_z=0.5t_0$.  With increasing $\lambda_z$, the spin splitting is enlarged and the change of energy of the bottom-right corner states (C$_1$ and C$_4$) is much faster than the other corner states (C$_2$ and C$_3$). When $\lambda_z$ is further increased, the corner states C$_1$ and C$_4$ merge into the edge states, and C$_2$ and C$_3$ switch their occupation statuses. Correspondingly, the edge states adjacent to the corners become more and more localized and generate new corner states (C$_5$ and C$_6$) in a concave shape, as the wavefunction distributions shown in Figure 2a for $\lambda_z=2t_0$. The evolution of energy levels in a larger regime can be found in Supporting Information Section II. The other two newly generated states (C$_7$ and C$_8$) are trivial localized states belonging to the corner hexagon on which $M$ is applied (see the two wavefunction distributions at the bottom-right of Figure 2a) so that the energy changing rate $\partial E/\partial \lambda_z$ is much larger than the corner states that contain some edge components. More details of the trivial localized states are discussed in the Supporting Information Section III.
   
 \begin{figure}[t]
 	\centering
 	\includegraphics[width=0.5\columnwidth]{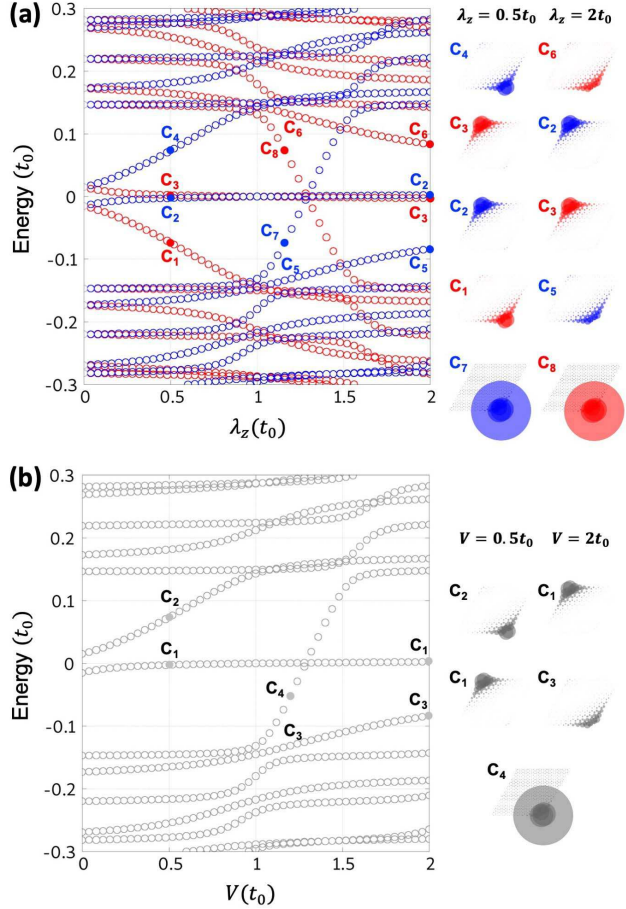}
 	\caption{\label{fig:epsart} Discrete energy levels of the corner states (a) as a function of local magnetism ($\lambda_z$) with $V=0$; (b) as a function of local electric potential ($V$) with $\lambda_z=0$. The red, blue, and grey circles represent the spin-up, spin-down, and spin-degenerated energy levels, respectively. The wavefunction distributions on the right panels correspond to the typical levels labeled as the solid circles on the left plots.}
 \end{figure}
 
 Next, we set $\lambda_z=0$ and focus on the influence of $V$. As shown in Figure 2b, the gap between bonding and antibonding corner states is enlarged with increasing $V$. The wavefunction of the corner states (C$_1$ and C$_2$) becomes unevenly distributed at the two corners with the addition of $V$, as shown in Figure 2b for $V=0.5t_0$. When $V$ increases, C$_2$ merges into the edge states. Meanwhile, a new corner state (C$_3$) emerges from the valence bands of edge states. The distribution of C$_3$ shows a concave shape, as shown in Figure 2b for $V=2t_0$. Again, a trivial localized state (C$_4$) forms at the bottom-right tip hexagon (see the bottom-right of Figure 2b). It is important to note that the persistence of having the corner states in our model against either $M$ or $V$ is another clear indication of its HOTI nature.
 
  \begin{figure}
 	\centering
 	\includegraphics[width=0.9\columnwidth]{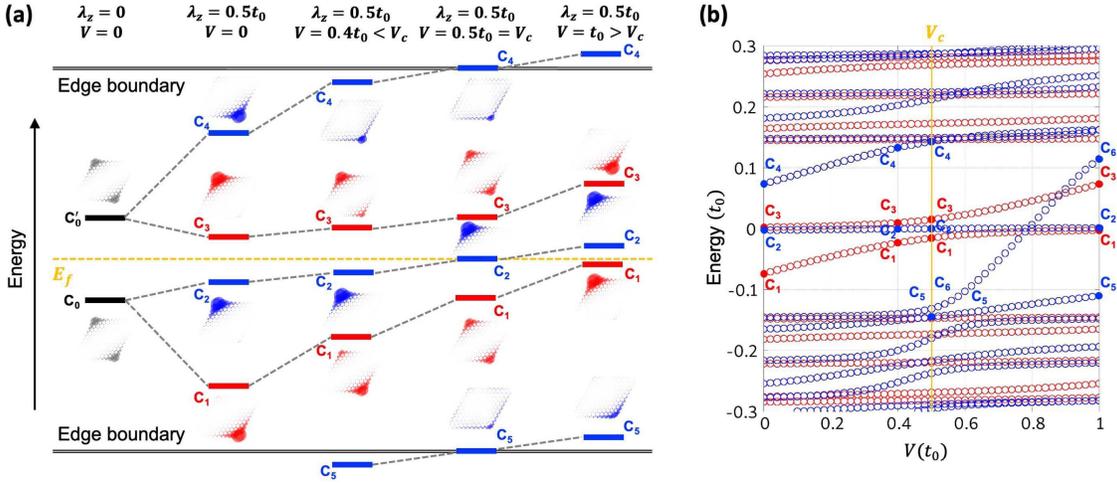}
 	\caption{\label{fig:epsart} Schematic figure illustrating the interplay of $M$ and $V$. The red, blue, and black dashes represent the spin-up, spin-down, and spin-degenerated energy levels, respectively. The inserted wavefunction distributions correspond to the levels labeled as the solid circles in (b). (b) Discrete energy levels of the corner states as a function of $V$ with $\lambda_z=0.5t_0$. The red and blue circles represent the spin-up and spin-down energy levels, respectively. $V_c$ is the SP inversion critical point.}
 \end{figure}
 
 Now, we investigate the response of the topological corner states to the interplay of $M$ and $V$. For simplicity, we plot the schematic figure of energy evolvements of corner states under the influence of $M$ and $V$ in Figure 3a. With $M$ is applied at the bottom-right corner, the spin degeneracy of corner states (C$_0$ and C$_0'$) is broken as already shown in Figure 2a. The spin-up bonding corner state (C$_1$) becomes more localized at the bottom-right corner, and spin-down bonding corner state (C$_2$) mostly locate at the top-left corner. Similar splitting and shift can also be found for the antibonding corner states (C$_3$ and C$_4$). If we further add a local positive $V$ at the same corner, all states at the bottom-right corner (C$_1$ and C$_4$) are lifted up, leading to the lifting of the corner states at the opposite corner (C$_2$ and C$_3$), as the evolvement of the energy levels shown in Figure 3b. When $V<V_c$, the energy difference between C$_1$ and C$_0$ (or C$_3$ and C$_0'$) decreases with increasing $V$. As a result, the wavefunctions of C$_1$ and C$_3$ weight more evenly between both corners. In contrast, the energies of C$_2$ and C$_4$ raise further with $V$ and become more localized at corners. Interestingly, C$_1$ and C$_3$ can be pulled back to their original states (C$_0$ and C$_0'$) at $V_c$, where $V_c=\lambda_z$, and restore the even distribution at two corners (see Figure 3a for $V=V_c$). When $V>V_c$, C$_2$ crosses the Fermi level and C$_4$ merges into the edge states. A new spin-down corner state (C$_5$) comes out from the top of edge bands. The spin-up corner state C$_1$ (C$_3$) becomes localized at the top-left (bottom-right) corner. The spin splitting of the two occupied corner states (C$_1$ and C$_5$) is reversed, which is effectively like a negative $M$ being applied on the bottom-right corner. If we compare the SP of the occupied corner states, its sign changes at both corners across $V<V_c$ and $V>V_c$, as shown in Figure 1(h,l).
 
 From the discussions above, there are three key messages from our systematic studies for the control of spin-polarized corner states: 1), magnetizing one corner can induce the SP of the other corner separated by a significant distance; 2), raising electric potential of one corner can differ the wavefunction distributions of the bonding and antibonding corner states; 3), the manipulation of $V$ with $M$ on the same corner can reverse the sign of SP at both corners. The use of negative $V$ shows the same conclusions (see Supporting Information Section IV). Besides, the shape or the size of the nanostructure is not constrained as the higher-order topology ensures the robustness of the corner states due to the bulk-boundary correspondence. A larger size of Kekul\'{e} nanostructure shows similar manifestations (see Supporting Information Section V). We also investigate the case having $M$ at both corners with ferromagnetic and antiferromagnetic orders. With $V$ applied on one corner, a transition happens between parallel and antiparallel SPs at two corners (see Supporting Information Section VI).
 
  \begin{figure}
 	\centering
 	\includegraphics[width=1\columnwidth]{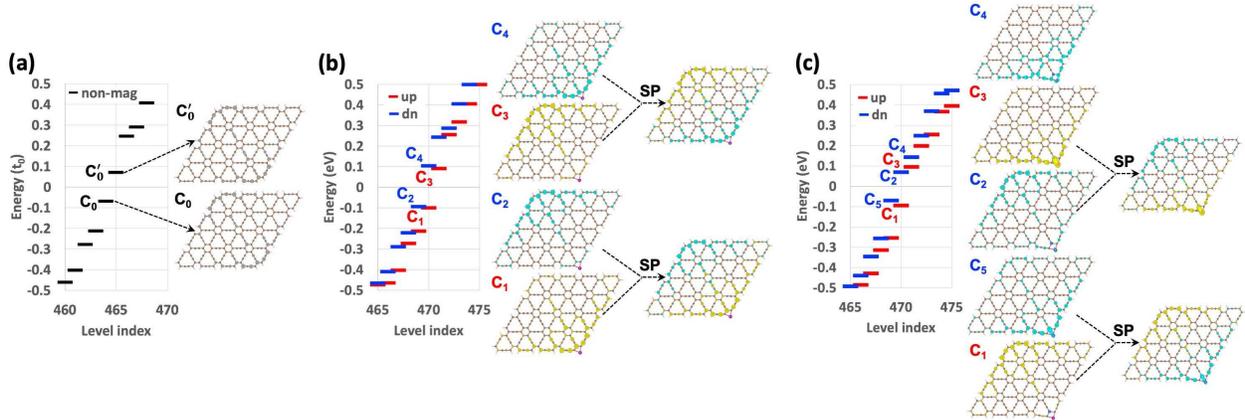}
 	\caption{\label{fig:epsart} Material realization of the $\gamma$-graphyne. Discrete energy levels of (a) 4$\times$4 $\gamma$-graphyne nanostructure (b) with Mn adsorption at the bottom-right corner (c) with Si dopant at the same corner, and the corresponding partial charge density for non-magnetic (grey), spin-up (yellow), and spin-down (blue) corner states (a) C$_0$ and C$_0'$ (b) C$_{1-4}$ (c) C$_{1-5}$. The spin polarization (SP) is calculated for the spin-polarized bonding and antibonding corner states in (b) and (c). Isovalue = 0.001/\AA.}
 \end{figure}

Last but not least, we discuss the possible realizations and advantages of our model using the first-principles calculations. The $\gamma$-graphyne nanostructures \cite{li2018synthesis,hu2022synthesis} are known as the HOTI due to the different intra-cell C-C and inter-cell C-C$\equiv$C-C bonds \cite{liu2019two}. We build a 4$\times$4 rhombus-shaped two-dimensional nanostructure, as shown in Figure 4a. The calculation method can be found in Supporting Information Section VII. Two corner states (C$_0$ and C$_0'$) within the edge gap are mainly distributed at the two 120$^{\circ}$ corners. In order to introduce $M$, we adsorb a Mn atom at the side of the bottom-right corner, which replaced two hydrogen adsorptions, as shown in Figure 4b. The system becomes spin-polarized with a magnetic momentum of 0.94$\mu_B$. The bonding (antibonding) corner state C$_0$ (C$_0'$) split into C$_1$ and C$_2$ (C$_3$ and C$_4$), which distribute at the bottom-right and top-left corners (top-left and bottom-right corners), respectively. The Mn atom adsorbed on the bottom-right corner induces the sizeable SP at both corners, as shown in the right panel of Figure 4b, in consistent with what is depicted in Figure 3 for $\lambda_z\neq0$ and $V=0$. Then, to mimic the addition of $V$ in DFT, we replace one C atom with Si to lift the on-site potential for both spin-up and spin-down states at the bottom-right corner, as shown in Figure 4c. All energy levels are lifted, the spin-up corner states (C$_1$ and C$_3$) become localized at the opposite corners, which is again comparable to the model results with $\lambda_z\neq0$  and $V>V_c$ in Figure 3. Spontaneously, a new spin-down corner state (C$_5$) is generated, similar to C$_5$ in Figure 3, except the alternation by Si makes C$_5$ higher than C$_1$. Interestingly, the sign of SP is reversed for both corners by $V$, as shown in the right panel of Figure 4c, which shows similar manifestations as we discussed in Figure 1(h,l).

To this end, we perceive that not only $\gamma$-graphyne but also any 2D HOTIs should manifest the topological features revealed in our model. The local magnetization can be applied by adding magnetic atoms near the corners, e.g., Mn, Cr, or Mo \cite{wu2009stability,wu2010electronic,thakur2016quest}; and the local electric potential can be easily controlled by the gate electrode\cite{yu2009tuning,wang2015electrically} or STM tips \cite{girard1993electric,yin2006nanoscale,ohara2008electric} in experiments.
 
In conclusion, we demonstrate that the local magnetization applied on one corner of a HOTI nanoflake can induce a sizeable SP at the other corners. Furthermore, manipulation of electric potential at the same corner may switch the sign of SP of both corners. Importantly, this remote control of topological corner states revealed from the TB model of Kekul\'{e} nanostructures is confirmed in a real material $\gamma$-graphyne through DFT calculations. The robust and remotely tunable spin-polarized corner states provide great advantages for diverse applications such as spin filtering, quantum gates, and information storage.

We thank Prof. Feng Liu at the University of Utah for helpful discussions. Work was supported by DOE-BES (Grant No. DE-FG02-05ER46237). Computer simulations were partially performed at the U.S. Department of Energy Supercomputer Facility (NERSC).

\providecommand{\latin}[1]{#1}
\makeatletter
\providecommand{\doi}
{\begingroup\let\do\@makeother\dospecials
	\catcode`\{=1 \catcode`\}=2 \doi@aux}
\providecommand{\doi@aux}[1]{\endgroup\texttt{#1}}
\makeatother
\providecommand*\mcitethebibliography{\thebibliography}
\csname @ifundefined\endcsname{endmcitethebibliography}
{\let\endmcitethebibliography\endthebibliography}{}

\pagebreak

\hfill \break
{\large Supporting Information for}
\\
\begin{center}
	\textbf{\large Remote control of spin polarization of topological corner states}
\end{center}

\hfill \break
\begin{center}
	{Yinong Zhou and Ruqian Wu$^{*}$}
\end{center}

\begin{center}
	{\textit{Department of Physics and Astronomy, University of California, Irvine, California 92697-4575, USA}}
\end{center}

\begin{center}
	{wur@uci.edu}
\end{center}

\clearpage

\sectionfont{\large}
\def\thesection{\Roman{section}}

\setcounter{equation}{0}
\setcounter{figure}{0}
\setcounter{table}{0}
\setcounter{page}{1}

\makeatletter
\renewcommand{\theequation}{S\arabic{equation}}
\renewcommand{\thefigure}{S\arabic{figure}}

\hfill \break

\tableofcontents

\clearpage

\section{Topological properties of Kekul\'{e} lattice}

\begin{figure}[h]
	\centering
	\includegraphics[width=0.6\columnwidth]{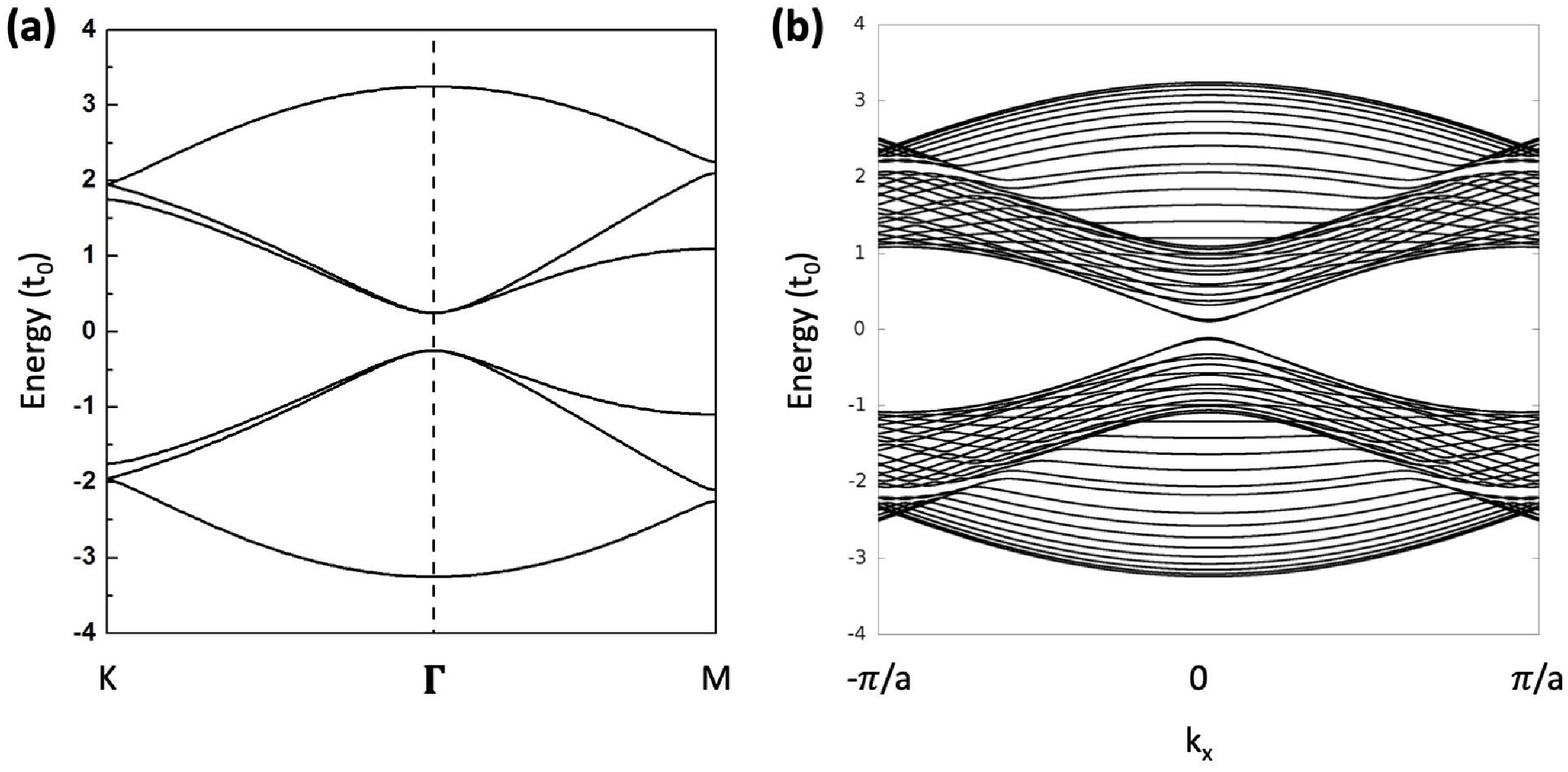}
	\caption{\label{fig:epsart} (a) The unit cell band structure. (b) The gapped edge states of the ribbon system. $t_1/t_0=1.25$, local magnetization ($M$) and local electric potential ($V$) equal to zero.}
\end{figure}

\renewcommand{\thetable}{S\arabic{table}}
\begin{table}[h]
	\caption{The parity at time-reversal invariant momenta, $\Gamma$ and M points for the six bands in Fig. S1a. The 1 to 6 band index represents the bands from bottom to top.}
	\begin{center}
		\includegraphics[width=0.8\columnwidth]{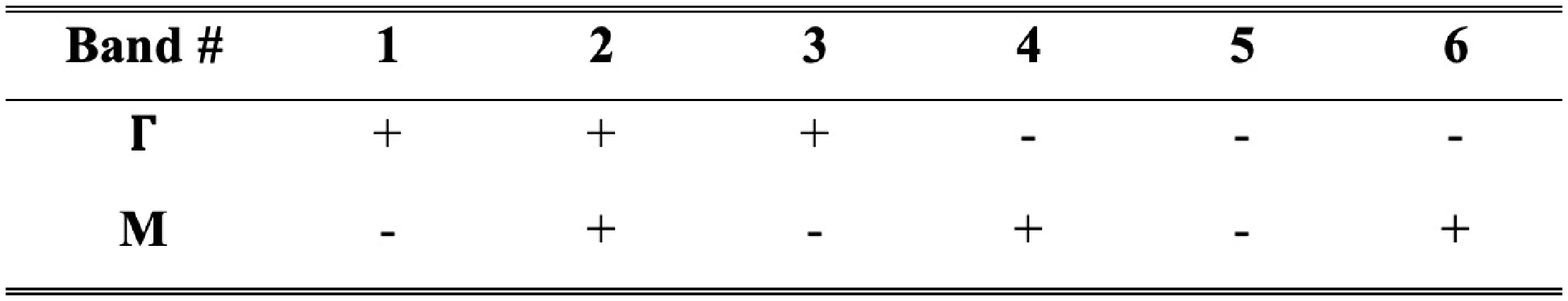}
	\end{center}
\end{table}

$Z_2$ number ($\nu$) can be calculated by using the parity eigenvalues at time-reversal invariant momenta \cite{hughes2011inversion,song2018diagnosis,fu2007topological,kim2015dirac,turner2012quantized,po2017complete}:
\begin{equation}
	{-1}^\nu=\left(-1\right)^{\left[N_{occ}^-\left(\Gamma\right)/2\right]}\times\left\{\left(-1\right)^{\left[N_{occ}^-\left(M\right)/2\right]}\right\}^3,
\end{equation}
where $N_{occ}^-$ is the number of the occupied bands with odd parity, and the square bracket represents the integer part of the number inside. The $Z_2$ number for the half occupation of Table S1 is calculated as $\nu=1$.

\clearpage

\section{Energy levels in a larger regime}

Here, we plot the evolvement of the energy levels with only $M$ or $V$ in a larger regime until $5t_0$ in Figure S2 as an extension of Figure 2 in the main text. As shown in Figure S2a, when $\lambda_z>t_0$, energy separation between spin-up and spin-down bonding (or antibonding) corner states decreases with further increasing $\lambda_z$. As shown in Figure 2a in the main text, the newly generated corner states (C$_5$ and C$_6$) are in a concave shape. Larger $\lambda_z$ weakens the coupling between the bottom-right corner hexagon and adjacent atoms so that the spin splitting decreases with further increasing $\lambda_z$. We note that except for the two trivial localized states C$_7$ and C$_8$ (labeled in Figure 2) generated from $\lambda_z=t_0$, there are two more trivial localized states generated from $\lambda_z=2t_0$. We will discuss more of these trivial localized states in the next section. Similar, in Figure S2b, with a larger $V$, the coupling between the bottom-right corner hexagon and adjacent atoms is weakened so that the energy separation between bonding and antibonding corner states decreases when $V>t_0$. And two trivial localized states are created from $\lambda_z=t_0$ and $\lambda_z=2t_0$, respectively. Then, they merge into the edge states with larger $V$.

Figure S3 shows the evolvement of the energy levels as a function of V when $\lambda_z=0.5t_0$ in a larger regime as an extension of Figure 3a in the main text. As we discussed in Figure 3a, a new spin-down corner state C$_5$ is generated when $V>\lambda_z$. With further increasing $V$, another new spin-up corner state C$_7$ (labeled in Figure S3) is created at the valence edge boundary. Meanwhile, the original corner state C$_1$ crosses the Fermi level. After C$_7$ is generated, the spin-up and spin-down bonding (or antibonding) states are localized at the same corner, see wavefunction distributions of C$_5$ and C$_7$ (or C$_1$ and C$_2$) in Figure S3. So, the regime to obtain opposite spin polarization (as shown in Figure 1k in the main text) is after the generation of C$_5$ is generated and before the generation of C$_7$.

\begin{figure}[h]
	\centering
	\includegraphics[width=0.7\columnwidth]{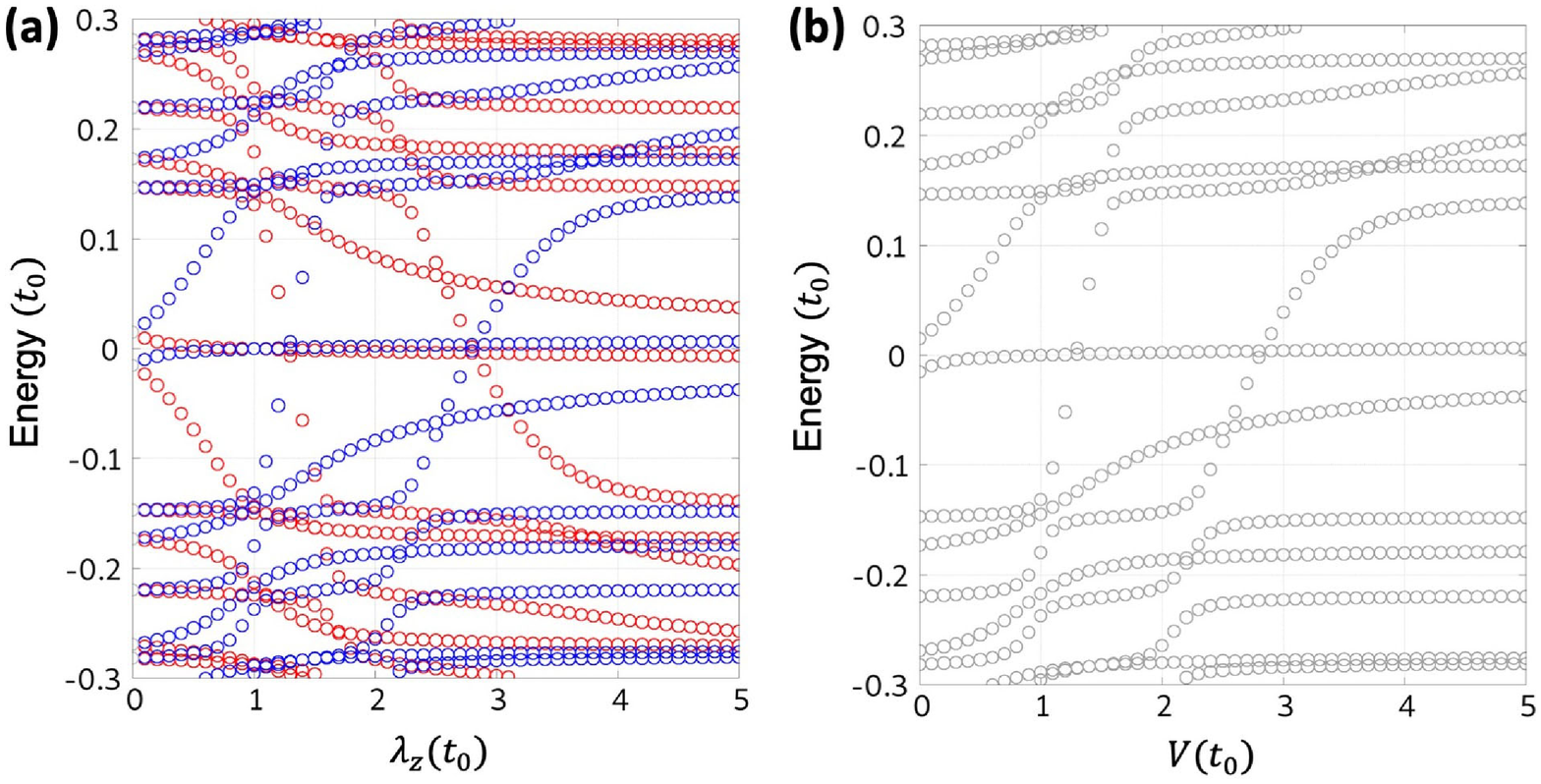}
	\caption{\label{fig:epsart} Discrete energy levels (a) as a function of $\lambda_z$ and $V=0$, (b) as a function of $V$ and $\lambda_z=0$ in a larger regime from 0 to $5t_0$. The red, blue, and grey circles represent the spin-up, spin-down, and spin-degenerated energy levels, respectively.}
\end{figure}

\begin{figure}[h]
	\centering
	\includegraphics[width=0.6\columnwidth]{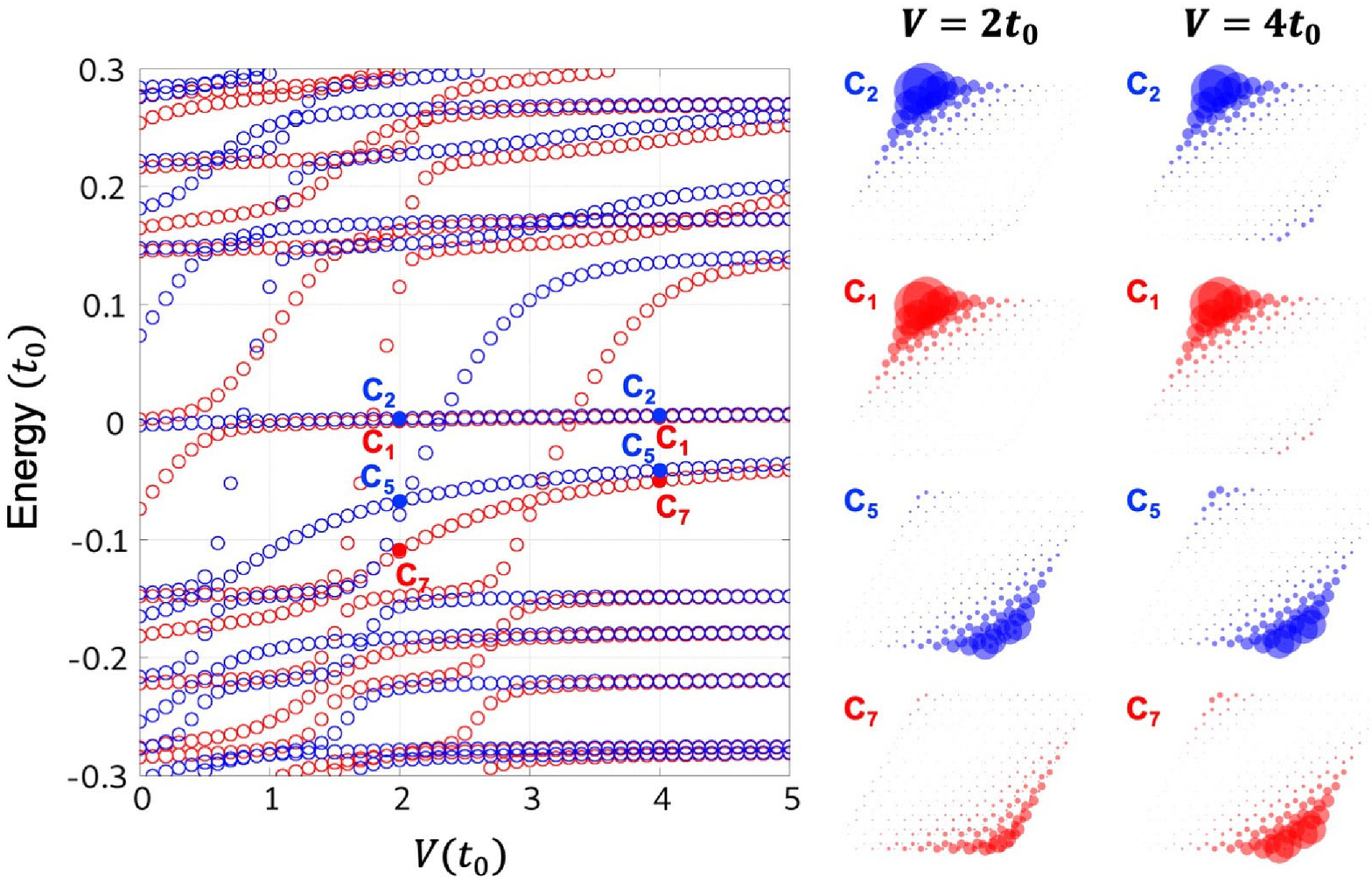}
	\caption{\label{fig:epsart} Discrete energy levels as a function of $V$ when $\lambda_z=0.5t_0$ in a larger regime from 0 to $5t_0$. The wavefunction distributions on the right panels correspond to the levels labeled as the solid circles on the left plots when $V=2t_0$ and $4t_0$.}
\end{figure}
\clearpage

\section{Trivial localized states}

Now, we investigate the trivial localized states that appear when we apply local $M$ or $V$. First, we discuss the trivial localized states generated with $\lambda_z=0.5t_0$ and increasing $V$ in a larger regime as shown in Figure S3. Four states (C$_6$, C$_8$, C$_9$, and C$_{10}$) are generated in sequence from the valence edge boundary and rapidly merge into the conduction edge boundary, as shown in Figure S4. These four states belong to the bottom-right corner hexagon, where we applied $M$ and $V$. We note that the inner and outer atoms of the hexagon have different surroundings that the inner three atoms are bonded with other atoms and the outer three atoms have no neighbors. As a result, the corner states C$_6$ and C$_8$ (or C$_9$ and C$_{10}$) appear in pairs distributed at the inner (or outer) three atoms of the hexagon (see Figure S4). These localized states created by the potential applied at the corner hexagon are topological trivial. The trivial localized states also appear with only applying $M$ or $V$, as shown in Figure 2 in the main text. We replot the trivial localized states (C$_7$ and C$_8$ in Figure 2a, and C$_4$ in Figure 2b) in Figure S5, which are distributed at the outer atoms of the corner hexagon.

\begin{figure}[h]
	\centering
	\includegraphics[width=0.5\columnwidth]{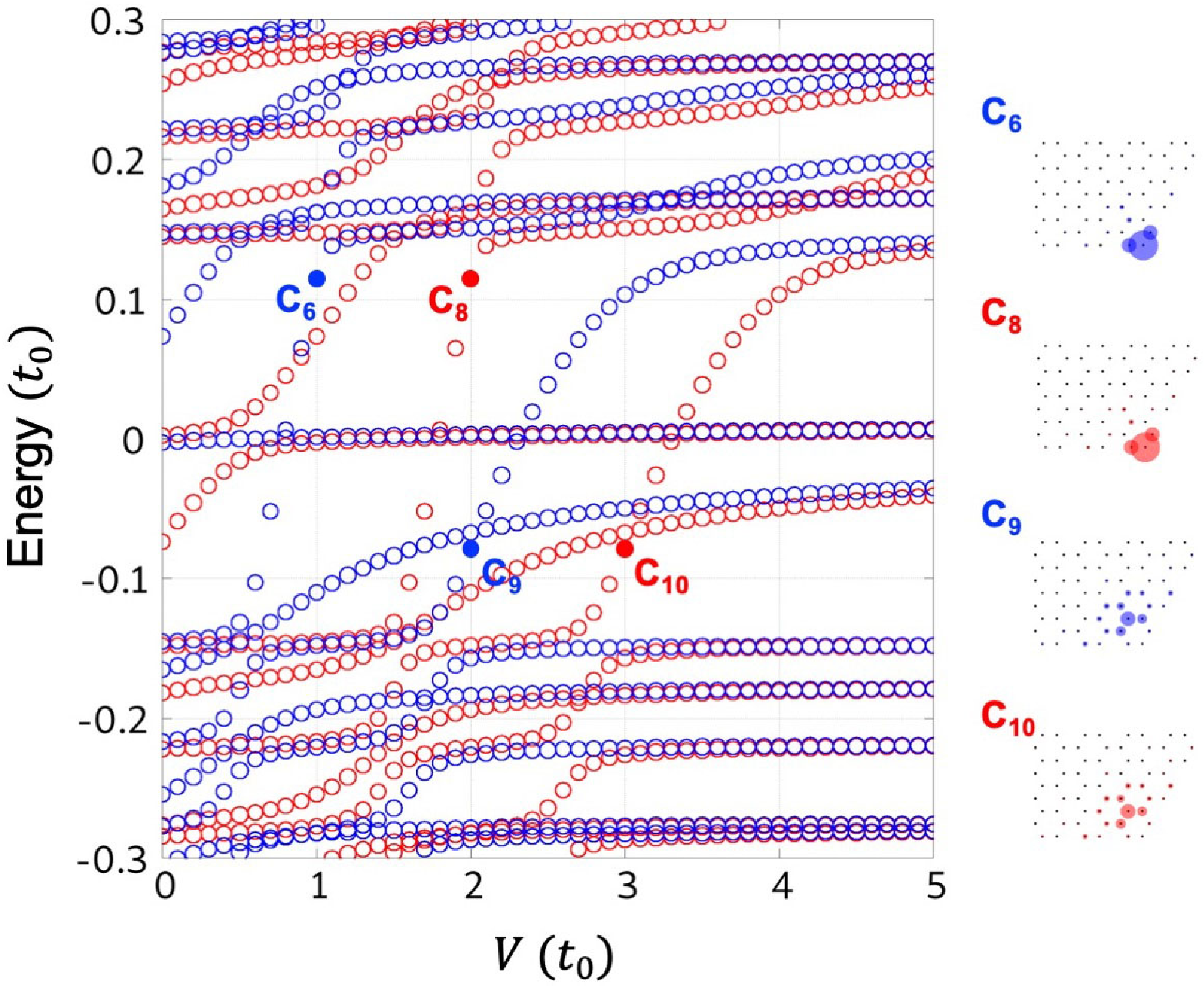}
	\caption{\label{fig:epsart} Four trivial localized states and their wavefunction distributions are plotted on the right. The size of the red and blue solid circles in the wavefunction distribution plots is proportional to the local charge density of corner states. To clarify the position of the circles, the size of the circles is decreased by 12.5 times compared to the wavefunction distribution in other figures.}
\end{figure}

\begin{figure}[h]
	\centering
	\includegraphics[width=0.6\columnwidth]{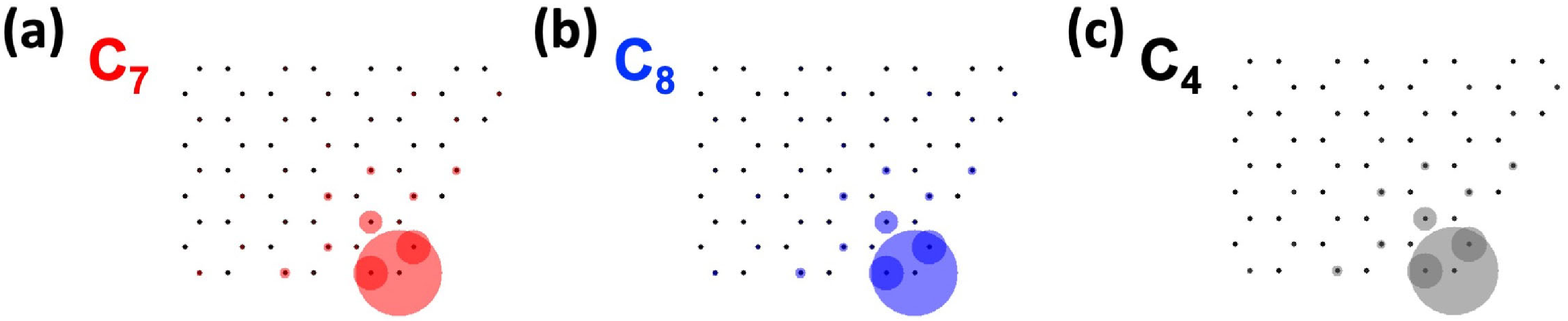}
	\caption{\label{fig:epsart} The zoom-in of the wavefunction distributions of the trivial localized states in Figure 2, (a) and (b) correspond to C$_7$ and C$_8$ in Figure 2a, (c) correspond to C$_4$ in Figure 2b. To clarify the position of the circles, the size of the circles is decreased by 12.5 times compared to the wavefunction distribution in Figure 2.}
\end{figure}
\clearpage

\section{Negative local electric potentials}

\begin{figure}[h]
	\centering
	\includegraphics[width=0.7\columnwidth]{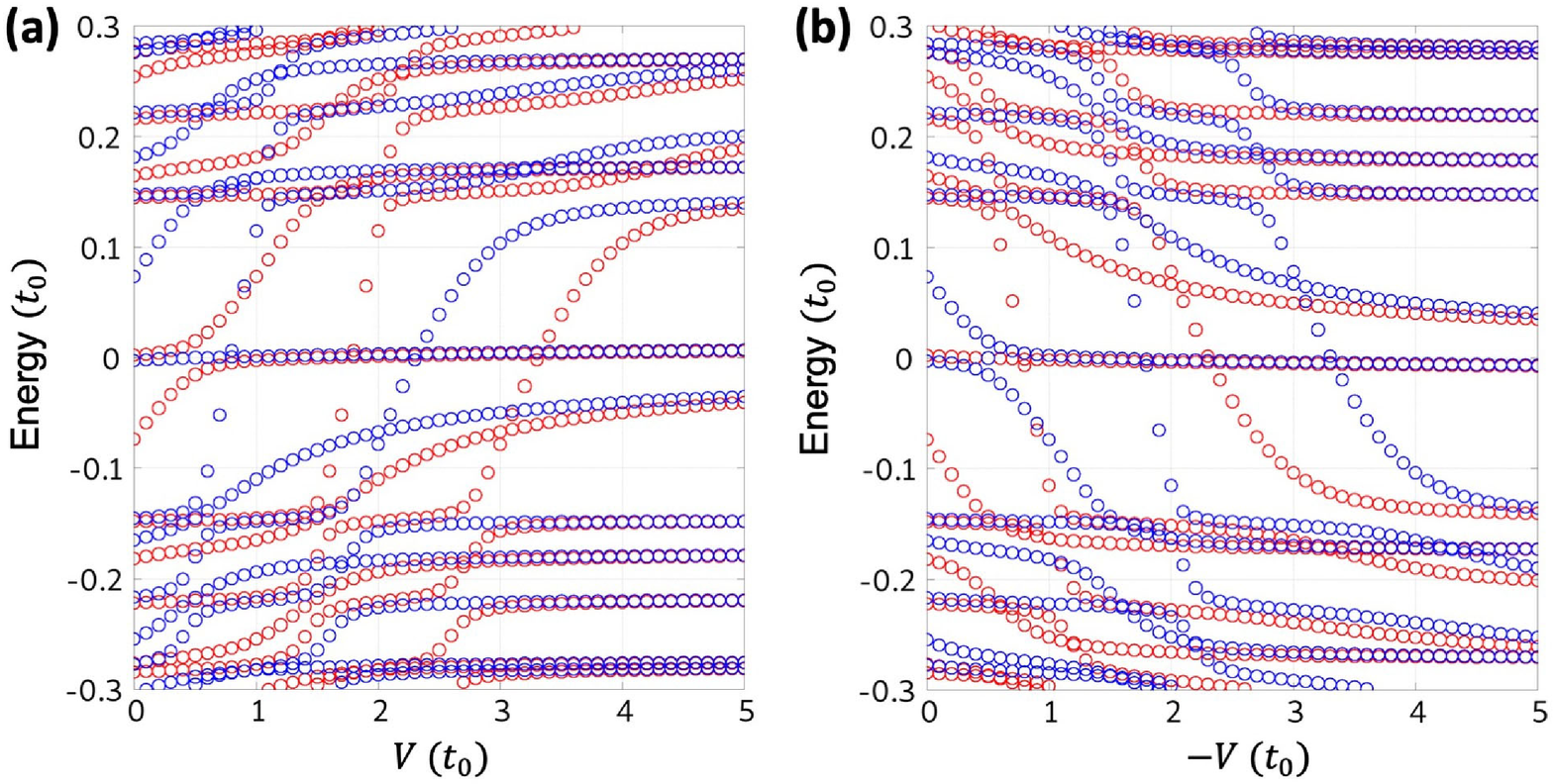}
	\caption{\label{fig:epsart} Discrete energy levels as a function of $V$ when $\lambda_z=0.5t_0$ for the (a) positive and (b) negative $V$. With opposite $V$, the evolvement of corner states is inverted for the bonding and antibonding states with the opposite spin.}
\end{figure}

\section{Larger sized nanostructures}

\begin{figure}[h]
	\centering
	\includegraphics[width=1\columnwidth]{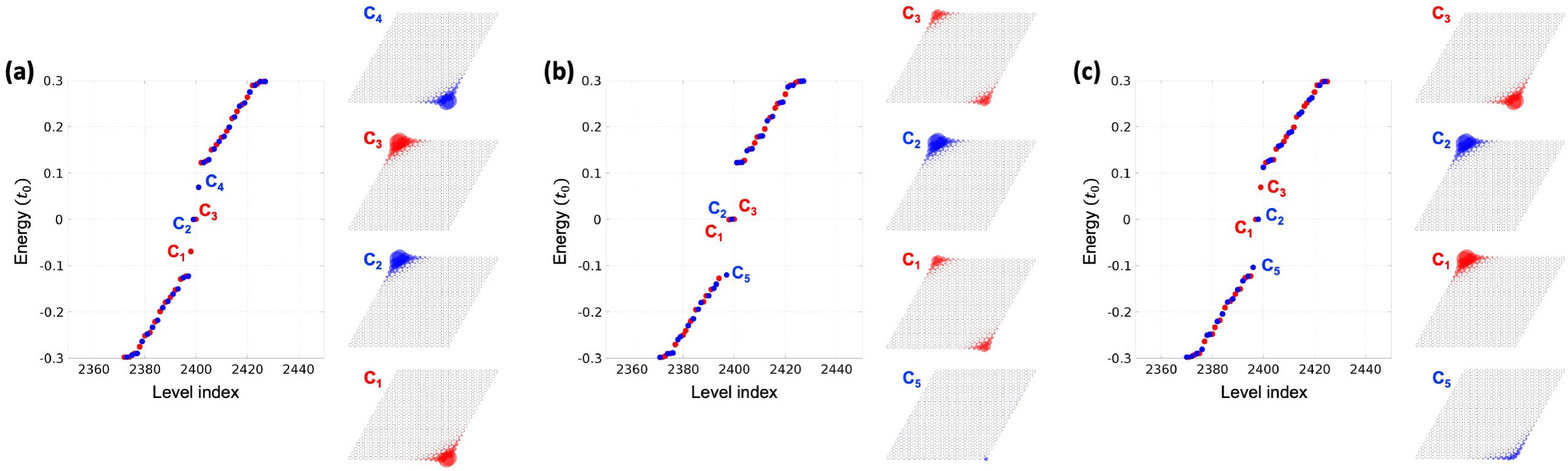}
	\caption{\label{fig:epsart} The discrete energy levels and the wavefunction distributions of corner states for a 20$\times$20 Kekul\'{e} nanostructure with $\lambda_z=0.5t_0$ and (a) $V=0$,  (b) $V=0.5t_0$, (a) $V=1t_0$, which shows similar results as the smaller sized nanostructure in Figure 3 in the main text.}
\end{figure}
\clearpage

\section{Model of local magnetization at both corners}

Next, we consider both corners are spin-polarized with ferromagnetic (FM) and antiferromagnetic (AFM) orders. As shown in Figure S8a, we applied $M$ on both corners with both positive magnetization (FM), bottom-right corner positive and top-left negative magnetization (AFM 1), and bottom-right corner negative and top-left positive magnetization (AFM 2). The spin polarization of different couplings is plotted in Figure S8b-d. Then, we apply $V$ on the bottom-right corner with $V>\lambda_z$, as shown in Figure S8e. The parallel coupling becomes antiparallel coupling for FM order, comparing Figure S8b and f. In contrast, the antiparallel coupling becomes parallel coupling for AFM orders, comparing Figure S8c,d and g,h. This transition between parallel and antiparallel couplings is due to the newly generated corner states at the corner applied $V$. The new corner states have opposite spin polarization as the original corner states, comparing spin polarization at the bottom-right corners in Figure S8b-d and f-h.

The evolvement of the energy levels as a function of $\lambda_z$ with $M$ on both corners when $V=0$ is shown in Figure S9, which can be seen as two copies of Figure S2a with $M$ applied on one corner. The spin degeneracy is broken for FM orders but is kept for AFM orders. Then, we apply $V$ on one corner. The evolvement of the energy levels as a function of $V$ when $\lambda_z=0.5t_0$ is shown in Figure S10. The new corner states are created when $V>\lambda_z$ for all three orders. And if we compare the total energy of FM and AFM orders (AFM 1 and AFM 2 have the same total energy), where the total energy is calculated as the summation of all occupied states. As shown in Figure S11, when $V<\lambda_z$, the AFM order is more stable as it has smaller total energy than the FM order. When $V>\lambda_z$, the SP of FM order transit from parallel (Figure S8b) to antiparallel (Figure S8f) at two corners, in contrast, the SP of AFM order transit from antiparallel (Figure S8c,d) to parallel (Figure S8g,h) at two corners. Since the antiparallel SP has lower total energy, the initial FM order becomes more stable than the AFM order.

\begin{figure}[h]
	\centering
	\includegraphics[width=0.8\columnwidth]{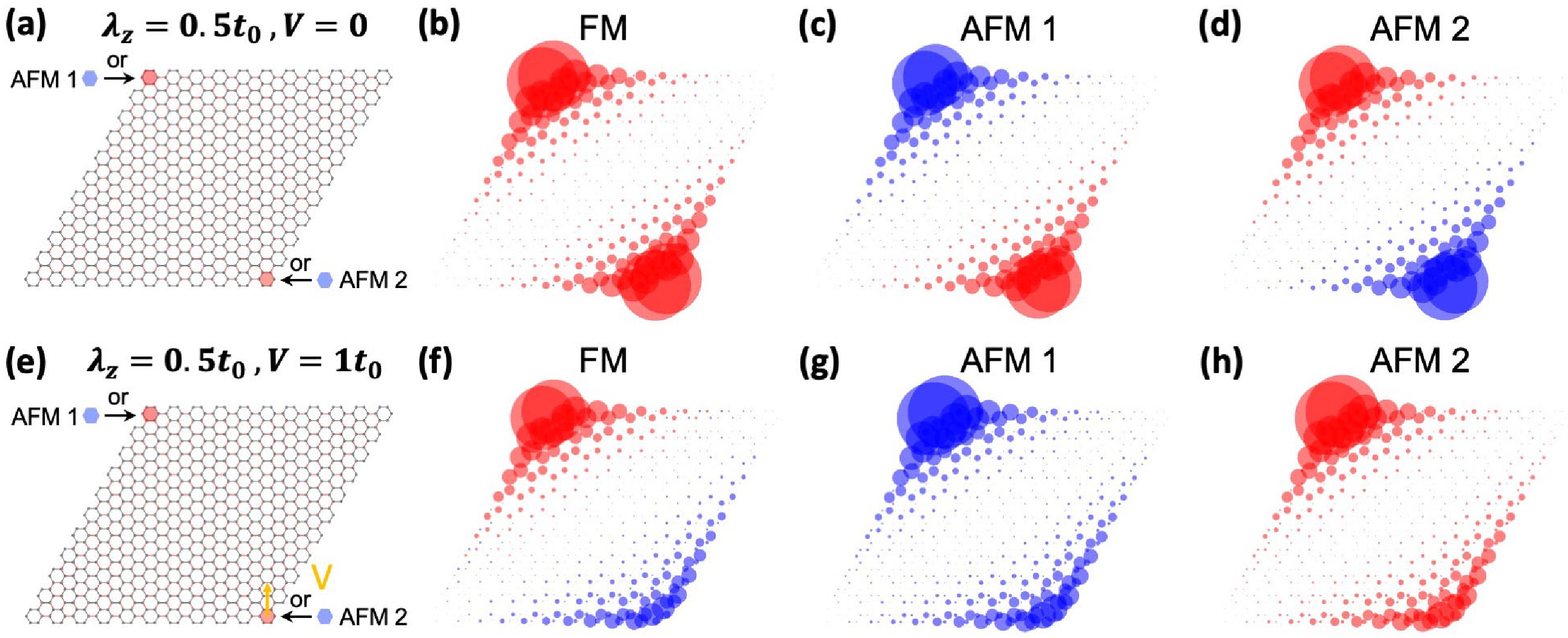}
	\caption{\label{fig:epsart} The rhombus-shaped Kekulé nanostructures (a) with $M$ at both corners with FM and AFM orders and (e) with $V$ at the bottom-right corner. The red and blue hexagons at the corners represent positive and negative $M$, respectively. Spin polarization distributions of the bonding corner states (b-d) without $V$ and (f-h) with $V$ for (b,f) FM, (c,g) AFM 1, and (d,h) AFM 2 orders. The red and blue circles are proportional to the local charge density of corner states.}
\end{figure}

\begin{figure}[h]
	\centering
	\includegraphics[width=0.35\columnwidth]{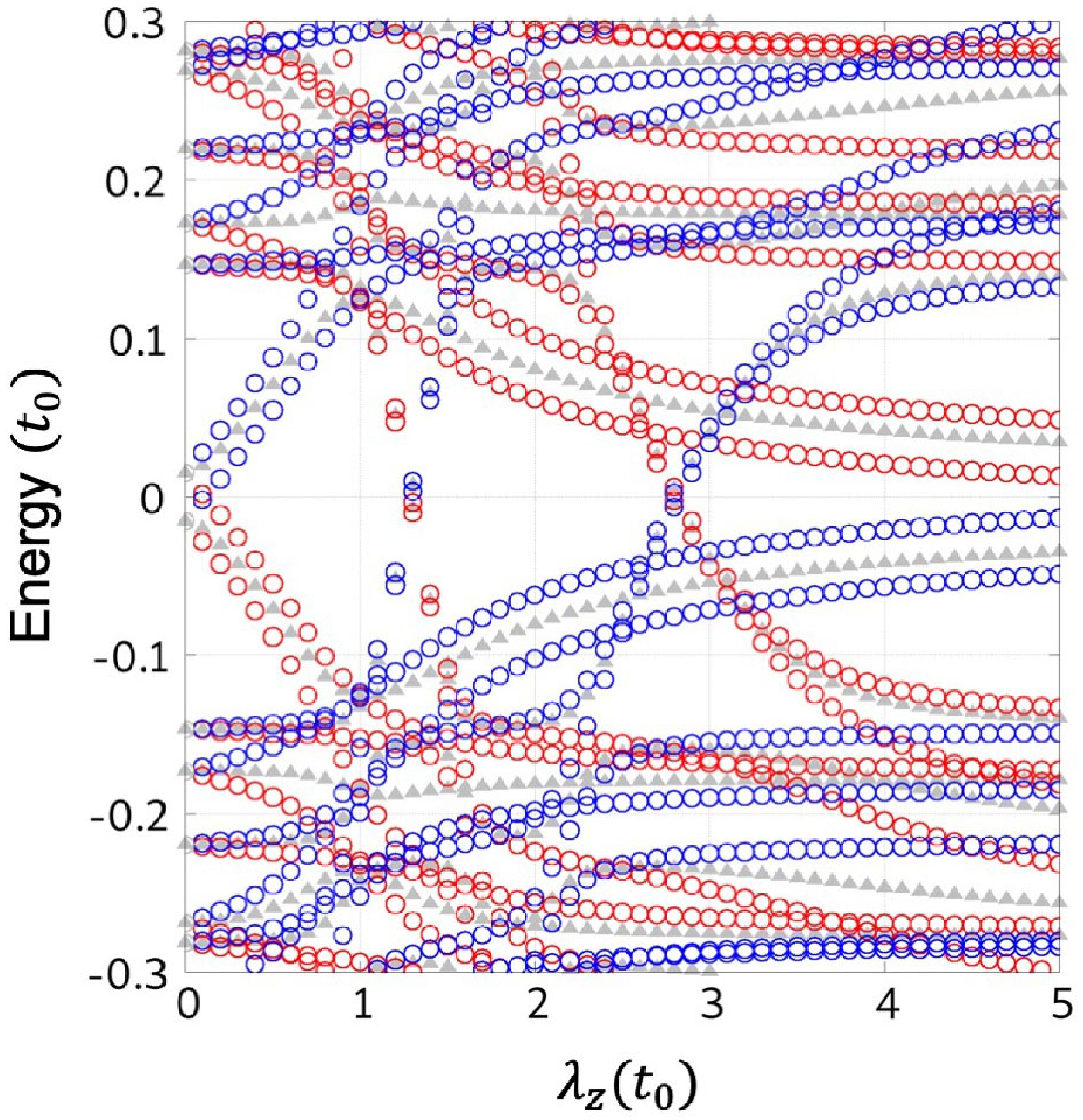}
	\caption{\label{fig:epsart} Discrete energy levels of FM and AFM orders as a function of $\lambda_z$ with $V=0$. The red (spin-up) and blue (spin-down) circles represent the energy levels with FM order, and the grey (spin degenerate) triangles represent the energy levels with AFM order.}
\end{figure}

\begin{figure}[h]
	\centering
	\includegraphics[width=1\columnwidth]{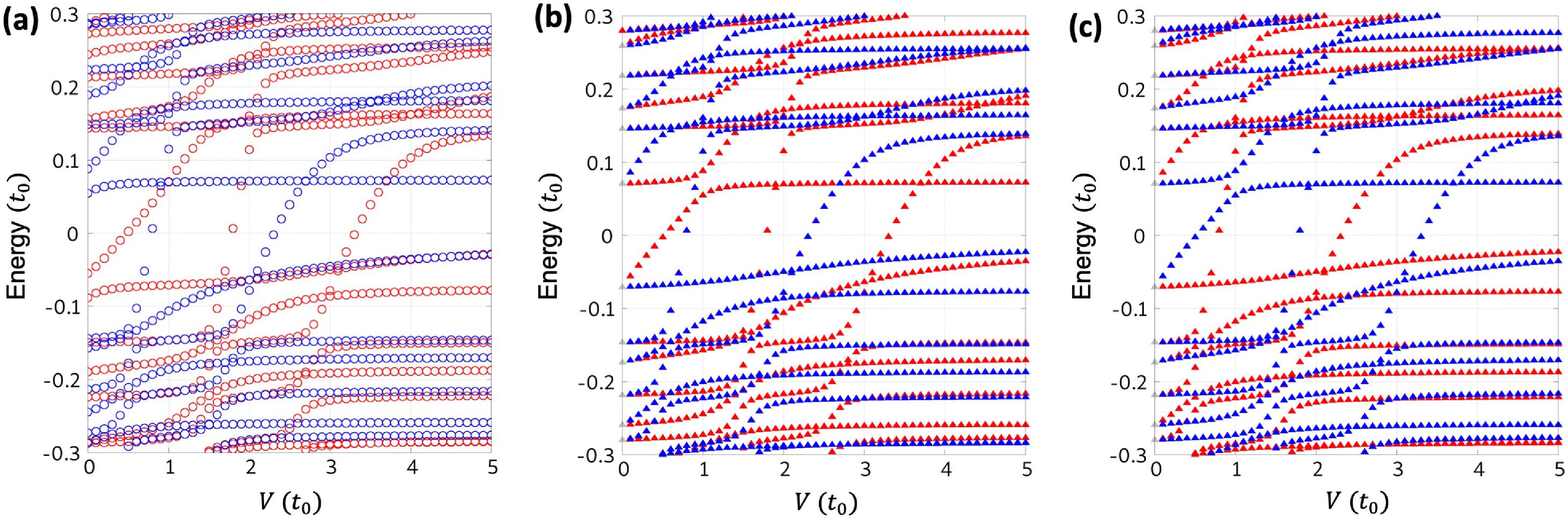}
	\caption{\label{fig:epsart} Discrete energy levels of (a) FM, (b) AFM 1 and (c)AFM 2 orders as a function of V with $\lambda_z=0.5t_0$. The red (spin-up) and blue (spin-down) circles/triangles represent the energy levels with FM/AFM orders.}
\end{figure}

\begin{figure}[h]
	\centering
	\includegraphics[width=0.5\columnwidth]{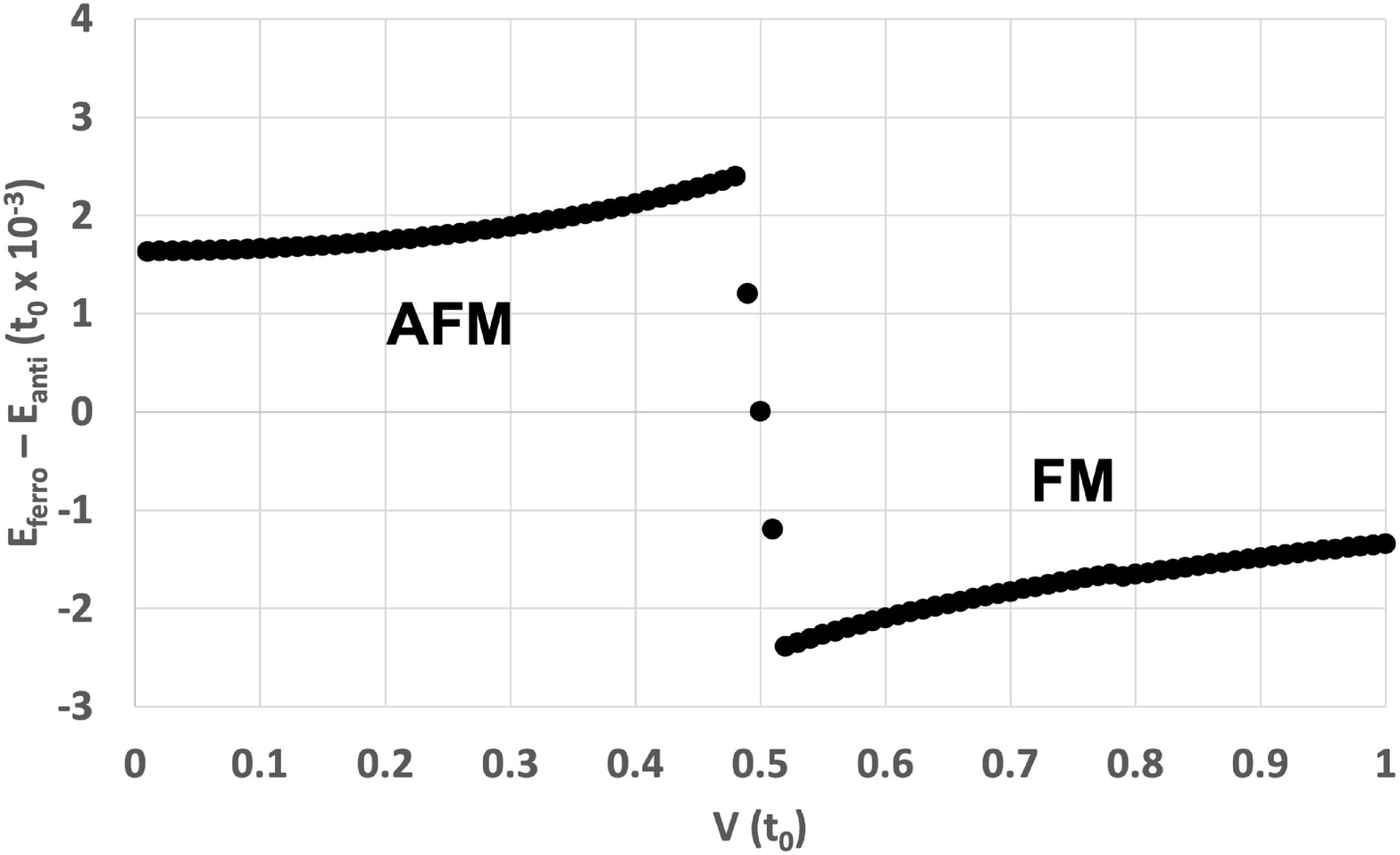}
	\caption{\label{fig:epsart} The energy difference between the FM and AFM orders as a function of $V$.}
\end{figure}

\clearpage

\section{\space First-principles calculation method}

Our first-principles calculations are performed with the projector-augmented wave pseudopotentials \cite{blochl1994projector,kresse1999ultrasoft} and the generalized gradient approximation of Perdew-Burke-Ernzerhof \cite{perdew1996generalized} using Vienna Ab initio Simulation Package \cite{kresse1996efficient} code. An energy cutoff of 450 eV and a 1$\times$1$\times$1 Monkhorst-Pack $k$-point grid is used \cite{methfessel1989high}. The structure is optimized until the atomic forces are smaller than 0.03 eV/\AA. The vacuum layer is larger than 15 \AA $\space$ to ensure decoupling between neighboring nanostructures. The magnetic moments are considered in the self-consistent calculation \cite{hobbs2000fully}. The DFT+U calculations are performed for the Mn-adsorbed $\gamma$-graphyne with $U = 4$ eV.

\clearpage

\providecommand{\latin}[1]{#1}
\makeatletter
\providecommand{\doi}
{\begingroup\let\do\@makeother\dospecials
	\catcode`\{=1 \catcode`\}=2 \doi@aux}
\providecommand{\doi@aux}[1]{\endgroup\texttt{#1}}
\makeatother
\providecommand*\mcitethebibliography{\thebibliography}
\csname @ifundefined\endcsname{endmcitethebibliography}
{\let\endmcitethebibliography\endthebibliography}{}

\end{document}